# Elastic, piezoelectric coefficients, and internal frictions of a single alpha-quartz crystal determined by partial-electrode electromechanical impedance spectroscopy

Mingyu Xie[1,2], Wenshuo Bai[1,2], Faxin Li[1,2,a)]

[1]LTCS and Department of Mechanics and Engineering Science, College of Engineering, Peking University, Beijing, 100871, China

[2] Center for Applied Physics and Technology, Peking University, Beijing 100871, China

**Abstract**

In this work, all the independent elastic coefficients, piezoelectric coefficients, and internal frictions of a single alpha-quartz crystal are determined using our recently proposed partial-electrode electromechanical impedance spectroscopy (PE-EMIS) at 25℃. In PE-EMIS, the rectangular parallelepiped quartz sample with two small partial electrodes fabricated on a corner is self-excited/sensed. The conductance spectrum (equivalent to the resonance spectrum) measured by an impedance analyzer under a 'true' free boundary condition is noiseless, allowing the first 100 eigenmodes ranging from 50 to 310kHz to be accurately fitted. To avoid mode misidentification, the off-plane displacement distributions of the quartz sample under different eigenmodes are determined using a scanned-laser vibrometer. The resonance spectrum measured using a commercial sandwich-like RUS apparatus is also presented for comparison, and the results show that the clamping force in sandwich-like RUS shifts the sample's resonance frequencies, causing 13% and 75% overestimations for the piezoelectric coefficients $e_{11}$ and $e_{14}$, respectively. In comparison to the RUS, our proposed PE-EMIS is more effective and convenient, and will be widely used for characterization of piezoelectric crystals.

**Keywords**: quartz crystal; elastic coefficients; piezoelectric coefficients; internal frictions; resonance spectrum; electromechanical impedance

---

a) Author to whom all correspondence should be addressed, Email: lifaxin@pku.edu.cn

# 1. Introduction

Since the Curie brothers discovered the piezoelectric effect in 1880, quartz crystal has been widely used as sensors,[1, 2] resonators,[3, 4] and wave filters[5, 6] due to its excellent mechanical, electrical, and temperature characteristics, as well as its extremely high Q value. Below 573°C, quartz is in alpha phase and possesses 32-point-group symmetry.[7] Alpha-quartz has six independent elastic coefficients, two independent piezoelectric coefficients, and two independent dielectric coefficients and can be written as follows:

$$c_{ij}^E = \begin{bmatrix} c_{11}^E & c_{12}^E & c_{13}^E & c_{14}^E & 0 & 0 \\ c_{12}^E & c_{11}^E & c_{13}^E & -c_{14} & 0 & 0 \\ c_{13}^E & c_{13}^E & c_{33}^E & 0 & 0 & 0 \\ c_{14}^E & -c_{14}^E & 0 & c_{44}^E & 0 & 0 \\ 0 & 0 & 0 & 0 & c_{44}^E & c_{14}^E \\ 0 & 0 & 0 & 0 & c_{14}^E & c_{66}^E \end{bmatrix} \quad (1)$$

Here, $c_{11}^E = 2c_{66}^E + c_{12}^E$. And,

$$e_{ij} = \begin{bmatrix} e_{11} & -e_{11} & 0 & e_{14} & 0 & 0 \\ 0 & 0 & 0 & 0 & -e_{14} & -e_{11} \\ 0 & 0 & 0 & 0 & 0 & 0 \end{bmatrix} \quad (2)$$

And

$$\kappa_{ij}^\varepsilon = \begin{bmatrix} \kappa_{11}^\varepsilon & 0 & 0 \\ 0 & \kappa_{11}^\varepsilon & 0 \\ 0 & 0 & \kappa_{33}^\varepsilon \end{bmatrix} \quad (3)$$

Accurately determining all the independent material coefficients, as well as the internal frictions of alpha-quartz are quite important for design of acoustic devices and has been reported by several researchers using different methods.[8-10]

In 1958, Bechmann[8] determined the complete set of elastic and piezoelectric coefficients of alpha-quartz based on the resonance of four differently-oriented bars and three differently-oriented plates. In the same year, Koga *et al.*[9] obtained these material coefficients according to the thickness vibration of thin rectangular quartz plates with different cut types. Kushibiki *et al.*[10] measured the longitudinal and shear wave velocity of three principal X-, Y-, and Z-cut quartz samples, as well as several rotated Y-cut samples to obtain all the independent matrix coefficients.

It can be seen that whether using the traditional resonance method or the pulse-echo method, multiple samples with specific shapes and orientations must be used. However, due to possible dimensional errors, crystal misorientation, and uncertain electrical boundary conditions of the multiple samples, the results obtained by these old methods maybe somewhat self-contradictory.[11] Furthermore, the internal frictions measurement by using these methods has never been reported.

In the past few decades, a new method called resonant ultrasound spectroscopy (RUS) has been developed to extract all the independent elastic coefficients of a single sample by matching the numerically computed eigenfrequencies to the measured acoustic resonance spectrum.[12-15] In RUS, the sample must be prepared with a well-defined shape, such as rectangular parallelepiped, sphere, cylinder, etc. In 1990, Ohno[16] firstly extended the RUS method to determine the elastic and piezoelectric coefficients of a piezoelectric crystal (alpha-quartz) by considering the piezoelectric effect in the eigenfrequencies calculation. In traditional RUS, the sample is sandwiched between two PZT transducers, one of which serves as a transmitter to generates elastic waves with constant amplitude and varying frequency, and the other as a receiver to detect vibration. The sandwich-like RUS method has a relatively strong signal due to the clamping force between the PZT transducers and the sample, but this force can cause the measured resonance frequencies differ from those obtained under free boundary condition.[17, 18] Ohno investigated the influence of this clamping force, and found that a 2g·wt force (the low limit for sandwich-like RUS) applied to a $3.219 \times 2.913 \times 3.690$ mm$^3$ alpha-quartz sample can cause a 0.2% increase in the resonance frequencies.[16] This clamping force also has a large influence on the damping measurement, especially for these expansion eigenmodes whose resonance energy can be easily transmitted into the transducers.[19]

To eliminate the influence of the clamping force, Ogi *et al.* adopted three transducers that form a tripod to determine the full matrix coefficients of alpha-quartz.[20] In the tripod-RUS, one of the transducers is a transmitter and the other two are receivers. When the sample is placed on the transducer tripod, the force is only the sample's weight, which is negligible for most practical sample sizes. In addition, Ogi *et al.*[21] and Johnson *et al.*[22] employed solenoid coil to measure the resonance spectrum of the piezoelectric crystals. The solenoid coil can induce an oscillating electric field near the surface of the piezoelectric crystal and excites vibration via the inverse piezoelectric effect. After excitation, same coil receives vibration through piezoelectric effect. Although these two methods can obtain the samples' 'true' free resonance frequencies, their coupling is too weak, making it hard to obtain a good signal-to-noise ratio (SNR). It seems that we must make a tradeoff between the perfectly free boundary condition and the SNR, but in fact, we can have both of them.

In this work, we use our recently proposed partial-electrode electromechanical impedance spectroscopy (PE-EMIS)[23] to determine all the independent elastic coefficients, piezoelectric coefficients and internal frictions of a single alpha-quartz sample without using any transducers. In PE-EMIS, the 'true' free resonance frequencies can be obtained and the measurement signals are noiseless. First, the principle of the PE-EMIS and the process for extracting elastic and piezoelectric coefficients are introduced. Second, the experimental details and calculation methods for determining internal frictions are presented. Then, the resonance spectrum of the alpha-quartz is measured using PE-EMIS from 50 to 310kHz and compared with that measured by a commercial sandwich-like RUS apparatus (Quasar RUSpec). Furthermore, we employ a scanned-laser vibrometer to unambiguously identify the resonance modes. Finally, the extracted material coefficients are analyzed and compared with that measured by other scholars.

## 2. Sample

Rectangular parallelepiped alpha-quartz sample provided by HuiJing Electronic Ltd is used for this study. The size of the sample is 15.06 (∥a-axis)×30.10(∥b-axis)×20.03(∥c-axis) mm$^3$. The dimension error is $\pm0.005$ mm, and the orientation error is $\pm0.05$ degree. The measured mass is 24.03g, so the density is 2.647g/ mm$^3$, which is much close to that measured by Ohno and Ogi.[16, 20]

# 3. Principle and experimental setup of PE-EMIS for determining all independent elastic and piezoelectric coefficients

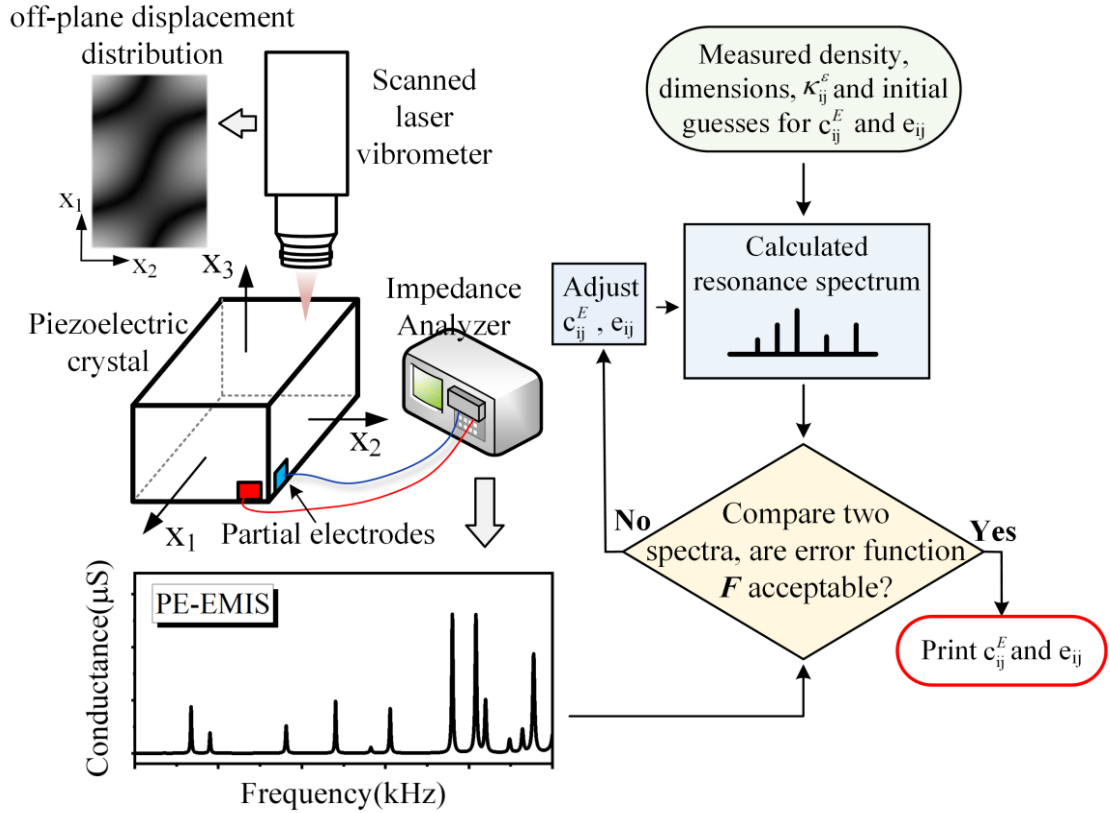

Fig.1 Principle and experimental setup of PE-EMIS for determining all independent elastic and piezoelectric coefficients.

To determine all independent elastic and piezoelectric coefficients of a single piezoelectric sample, we must accurately measure the free resonance acoustic spectrum, and usually, the number of measured resonance frequencies should be at least 5 times the number of the independent material coefficients to be determined.[14] However, due to the weak piezoelectric coefficients in alpha-quartz, their normalized sensitivities to the resonance frequencies are always less than 1%. So, to guarantee the weak piezoelectric coefficients to be accurately extracted, the number of measured resonance frequencies should be at least 10 times, preferably 12-14 times the number of independent material coefficients to be determined. Traditional sandwich-like RUS using two PZT transducers to clamp the sample so that the sample's resonance frequencies will be slightly shifted. Recently, we proposed a new method called partial-electrode electromechanical impedance spectroscopy (PE-EMIS) that can determine the 'true' free resonance frequencies of piezoelectric sample without the use of any transducers.[23] As shown in Fig.1, in PE-EMIS test, two 'smallest' partial electrodes are made on a corner of the alpha-quartz sample and the processes are as follows: i) Wrap the rectangular parallelepiped quartz sample in adhesive tape, and leave only two square areas for painting SPI conductive silver (~5μm thick) as electrodes. ii) Remove the adhesive tape and use the impedance analyzer to detect the sample's first three-order conductance curves when the electrodes are dried. iii) Remove the electrodes with acetone and make two smaller new electrodes again. iv) Repeat the above processes until the first three-order conductance signals are difficult to detect. In this way, we can find the "smallest" electrode area, which is about 4×4mm$^2$ for this quartz sample (15.06×30.10×20.03mm$^3$). The extraction error of material coefficients caused by the 'smallest' electrode is usually negligible,[23] and can be proved by finite element method (FEM). Fig.2 shows the FEM calculation for the influence of the sizes of the partial electrodes on the sample's first 100 free resonance frequencies, and it can be seen that these two 'smallest' electrodes (4×4mm$^2$) reduce the free resonance frequencies by less than 0.01%, which is negligible. After that, the conductance spectrum, equivalent to the free resonance acoustic spectrum, can be measured with a commercial impedance analyzer connected to these two partial electrodes through two soft thin wires, as shown in Fig.1. After finishing the measurement of complete resonance spectrum, a non-conductive reflective paint is sprayed on the $x_3$ face (15.06×30.10mm$^2$) of the quartz crystal with the thickness of about 1μm, and the off-plane displacement distributions of the sample under different resonance modes are then measured using a scanned-laser vibrometer.[20, 21] The measured off-plane displacement patterns, as the unique 'fingerprint' of each eigenmode, will be compared with the calculated one later to avoid mode misidentification.

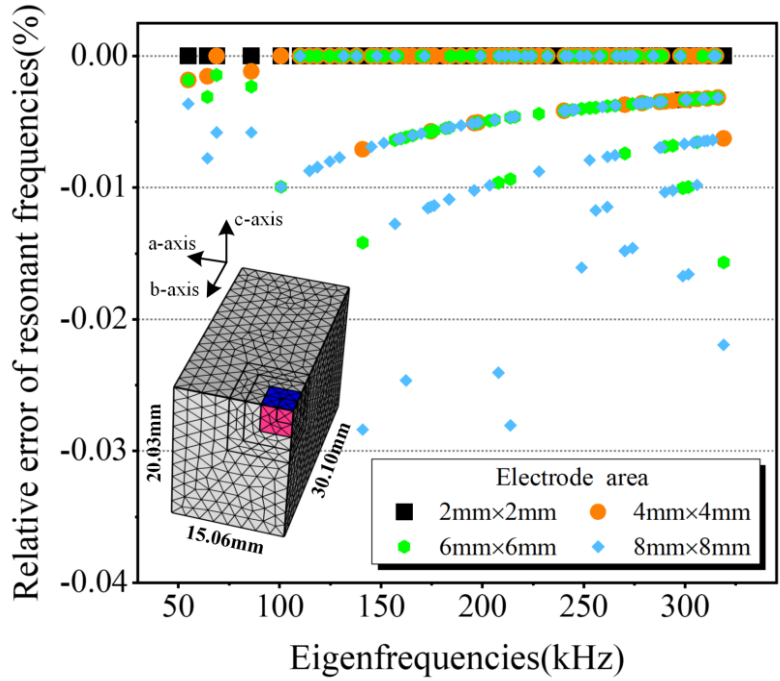

Fig.2 FEM analysis on the relative errors between the eigenfrequencies of electrodeless quartz sample and the resonance frequencies obtained from the conductance curve of the partial-electrode sample.

When the complete resonance spectrum has been measured, we can extract all the independent elastic and piezoelectric coefficients of the rectangular parallelepiped alpha-quartz by considering the forward problem and the backward problem, and the flow chart is shown in the right column of Fig. 1. The forward problem uses Lagrangian variation to calculate the resonance frequencies according to the density, dimensions, dielectric, elastic, and piezoelectric coefficients of the sample, while the backward problems employs L-M algorithm[24] to inverse all the independent elastic and piezoelectric coefficients. It should be noted that the dielectric coefficients cannot be directly inversed and should be determined ahead of time.[23] Here we directly used the values reported by Ogi *et al.*,[20] which are 4.424 for $\kappa_{11}^{\varepsilon}/\varepsilon_0$, and 4.632 for $\kappa_{33}^{\varepsilon}/\varepsilon_0$, respectively. Here, $\varepsilon_0$ is the vacuum-permittivity.

## A. Forward problem

The variation of Lagrangian function in piezoelectricity is given by:[16, 25]

$$\delta L = \delta \int_V \frac{1}{2} \left( \varepsilon_{ij} c^E_{ijkl} \varepsilon_{kl} - \phi_{,m} \kappa^\varepsilon_{mn} \phi_{,n} + 2\phi_{,m} e_{mkl} \varepsilon_{kl} - \rho \omega^2 u_i u_i \right) dV = 0 \qquad (4)$$

where $c^E_{ijkl}$, $\kappa^\varepsilon_{mn}$, and $e_{mkl}$ are elastic, dielectric, and piezoelectric coefficients, respectively. $\rho$ is the mass density, $\omega = 2\pi f$ is the angular frequency. $\varepsilon_{ij} = \frac{1}{2}(u_{i,j} + u_{j,i})$ is the strain. $u_i$ and $\phi$ and are displacement and electric potential and can be expressed as:

$$\mathbf{u}(x_1, x_2, x_3) = \sum_{p=1}^{N} a^{(p)} \boldsymbol{\varphi}^{(p)} \qquad (5)$$

$$\phi(x_1, x_2, x_3) = \sum_{q=1}^{M} b^{(q)} \psi^{(q)} \qquad (6)$$

where $a^{(p)}$ and $b^{(q)}$ are the corresponding expansion coefficients. Here, the normalized Legendre polynomial $\bar{P}_\lambda$ is used:

$$\boldsymbol{\varphi}^{(p)} = (l_1 l_2 l_3)^{-1/2} \bar{P}_\lambda(x_1/l_1) \bar{P}_\mu(x_2/l_2) \bar{P}_\nu(x_3/l_3) \mathbf{e}_n \qquad (7)$$

$$\psi^{(q)} = (l_1 l_2 l_3)^{-1/2} \bar{P}_\alpha(x_1/l_1) \bar{P}_\beta(x_2/l_2) \bar{P}_\gamma(x_3/l_3) \qquad (8)$$

where $x_i$ is the Cartesian coordinates (seen in Fig.1). $l_i$ is the half length of the $i^{th}$ dimension of the sample. λ, μ, and ν are the order number of Legendre polynomial. By substituting Eqs. (5) and (6) into Eq. (4), we have:

$$(\boldsymbol{\Gamma} + \boldsymbol{\Omega}^T \boldsymbol{\Lambda}^{-1} \boldsymbol{\Omega}) \mathbf{a} = \rho \omega^2 \mathbf{a} \qquad (9)$$

$$\mathbf{b} = \boldsymbol{\Lambda}^{-1} \boldsymbol{\Omega} \mathbf{a} \qquad (10)$$

Where $\boldsymbol{\Gamma}, \boldsymbol{\Omega}$ and $\boldsymbol{\Lambda}$ are elastic, piezoelectric and dielectric interaction matrices:

$$\Gamma_{pp'} = \int_V S_{ij}(\boldsymbol{\varphi}^{(p)}) c_{ijkl} S_{kl}(\boldsymbol{\varphi}^{(p')}) dV \qquad (11)$$

$$\Omega_{qp} = \int_V \psi^{(q)}_{,m} e_{mkl} S_{kl}(\boldsymbol{\varphi}^{(p)}) dV \qquad (12)$$

$$\Lambda_{qq'} = \int_V \psi^{(q)}_{,m} \kappa_{mn} \psi^{(q')}_{,n} dV \qquad (13)$$

$$S_{ij}(\boldsymbol{\varphi}^{(p)}) = \frac{1}{2}(\varphi^{(p)}_{i,j} + \varphi^{(p)}_{j,i}) \qquad (14)$$

For a rectangular parallelepiped piezoelectric crystal with the symmetry higher than 2/m, the resonance modes can be divided into four groups[16, 26] denoted by Ag, Au, Bg, Bu, and can be obtained by solving the eigenvalues of Eq.(9). This eigenmode classification not only helps mode identification but also can accelerate the numerical calculation.

## B. Backward problem

As shown in the right column of Fig. 1, the backward problem is using the inversion algorithm to optimize the material coefficients to achieve a minimizer of the following error function:

$$F = \sum_{k=1}^{Num} w_k \left( f_k^{cal} - f_k^{meas} \right)^2 \tag{15}$$

where $f_k^{cal}$ and $f_k^{meas}$ are the $k^{th}$ calculated and measured eigenfrequencies. $w_k$ is the weighting factor and is normally set as to $(f_k^{meas})^{-2}$.[12] However, when mode omission happens, we must manually add the resonance frequencies and set $w_k$ to be zero.[14] Here a self-written L-M algorithm is used and the iteration steps are as follows:

$$\Delta \mathbf{p} = -[\mathbf{A} + \lambda \mathbf{I}]^{-1} \mathbf{B} \tag{16}$$

$$A_{\alpha\beta} = \sum_{k=1}^{Num} 2w_k \frac{\partial f_k^{cal}}{\partial p_\alpha} \frac{\partial f_k^{cal}}{\partial p_\beta} \tag{17}$$

$$B_\beta = \sum_{k=1}^{Num} 2w_k \left( f_k^{cal} - f_k^{meas} \right) \frac{\partial f_k^{cal}}{\partial p_\beta} \tag{18}$$

where $\Delta \mathbf{p}$ is the increment of the elastic and piezoelectric coefficients during each iteration step. $\mathbf{I}$ is the identity matrix, and $\lambda$ is the damping parameter.

## 4. Principle and experimental setup of PE-EMIS for determining very low internal frictions of piezoelectric crystals

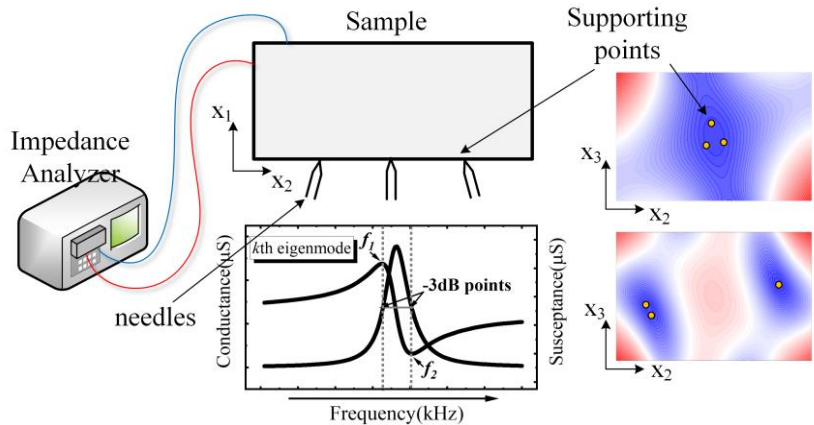

Fig.3 Principle and experimental setup of PE-EMIS for determining very low internal frictions of piezoelectric crystals.

As shown in Fig.1, for elastic and piezoelectric coefficients determining, the quartz sample can be supported by a soft foam with very low acoustic impedance, and the complete spectrum can then be conveniently detected using an impedance analyzer in one test. However, for the measurement of very low internal frictions (damping) in alpha-quartz sample, we must use sharp needles to support it near the displacement nodes of each eigenmode to avoid additional energy loss, as shown in Fig.3. The internal friction of $k^{th}$ eigenmode $Q_k^{-1}$ can then be determined one by one with the impedance analyzer, and the calculation formula is as follows:[27, 28]

$$Q_k^{-1} = \frac{\Delta f_k}{f_k} = 2\frac{f_2 - f_1}{f_2 + f_1} \tag{19}$$

where $f_1$ and $f_2$ are the two -3dB points of the conductance curve, and are also the resonance and antiresonance points of the susceptance curve.[27] $f_k = (f_1 + f_2)/2$ is the mechanical resonance frequency, and $\Delta f_k = f_2 - f_1$ is the half bandwidth. For a piezoelectric crystal, we have proved that the internal friction of $k^{th}$ eigenmode $Q_k^{-1}$ can be generally expressed as:[23]

$$\frac{1}{Q_k} = \frac{2}{f_k}\left(\sum_{ij}\frac{\partial f_k}{\partial c_{ij}^E}c_{ij}^E\frac{1}{Q_{ij}^c} - \sum_{pq}\frac{\partial f_k}{\partial \kappa_{pq}^\varepsilon}\kappa_{pq}^\varepsilon\frac{1}{Q_{pq}^\kappa} + \sum_{mn}\frac{\partial f_k}{\partial e_{mn}}e_{mn}\frac{1}{Q_{mn}^e}\right) \tag{20}$$

Where $\frac{\partial f_k}{\partial c_{ij}^E}$, $\frac{\partial f_k}{\partial \kappa_{pq}^\varepsilon}$ and $\frac{\partial f_k}{\partial e_{mn}}$ are the sensitivities of the elastic, dielectric and piezoelectric coefficients to the $k^{th}$ resonance frequency $f_k$. $\frac{1}{Q_{ij}^c}$, $\frac{1}{Q_{pq}^\kappa}$, and $\frac{1}{Q_{mn}^e}$ are the independent elastic, dielectric, and piezoelectric internal frictions, respectively. Determining all these independent internal frictions allows one to predict the loss of any vibration mode. For materials with strong piezoelectricity, such as lead zirconate titanate ferroelectric ceraimics, $\frac{1}{Q_{pq}^\kappa}$ and $\frac{1}{Q_{mn}^e}$ cannot be neglected. However, for alpha-quartz, the influence of dielectric and piezoelectric coefficients on the resonance frequencies are extremely low. So, the half power bandwidth of the resonance peak is dominated by $\frac{1}{Q_{ij}^c}$, and in this condition, Eq. (20) can then be simplified as:

$$\frac{1}{Q_k} = \frac{2}{f_k}\sum_{ij}\frac{\partial f_k}{\partial c_{ij}^E}c_{ij}^E\frac{1}{Q_{ij}^c} \tag{21}$$

## 5. Results and discussions

### 5.1 Elastic and piezoelectric coefficients

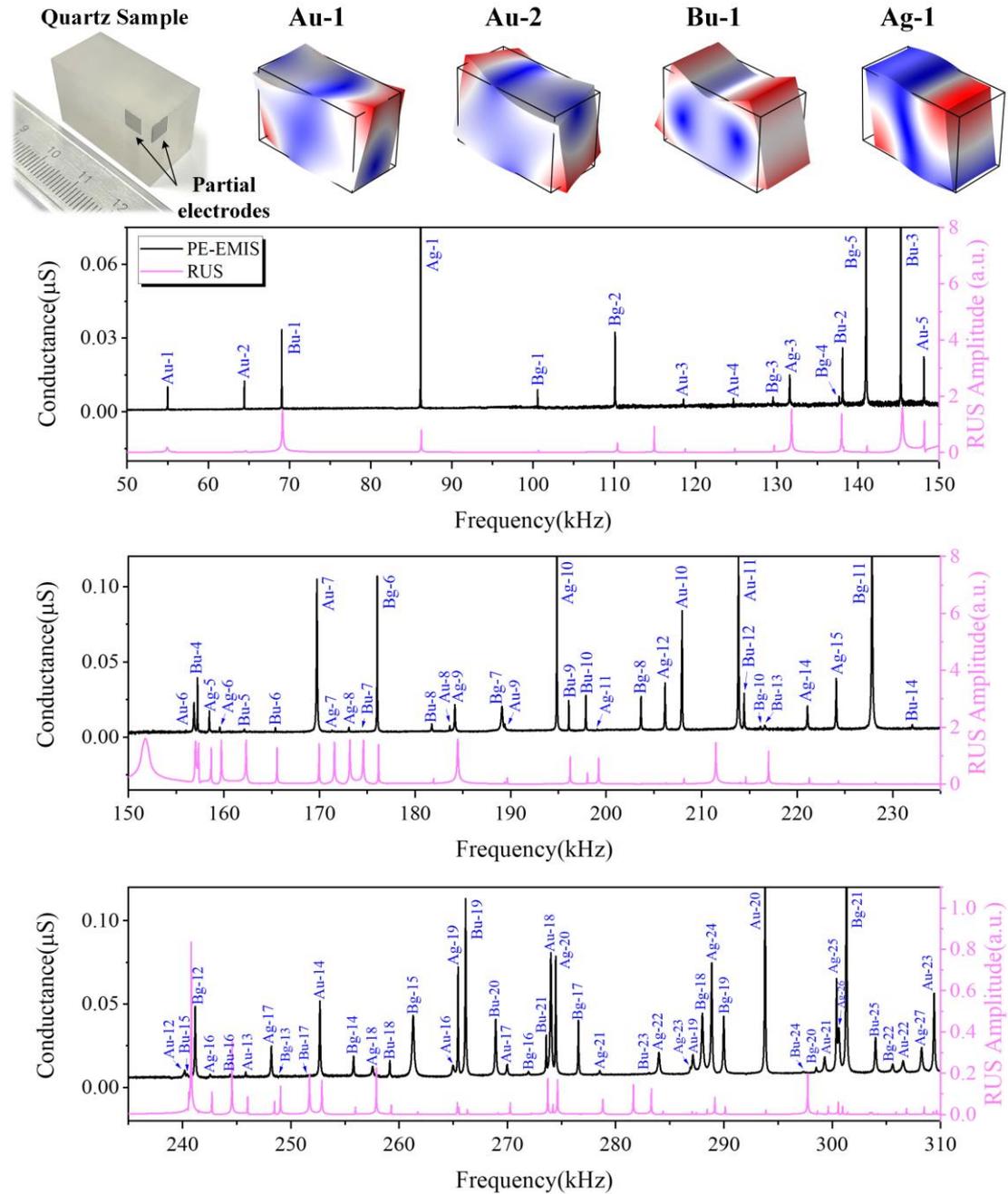

Fig.4 Acoustic resonance spectrum of rectangular parallelepiped alpha-quartz sample measured from 50 to 310kHz at 25℃. Black line: electromechanical impedance spectrum measured by PE-EMIS; Pink line: resonance spectrum measured by commercial sandwich-like RUS (Quasar RUSpec).

The black lines in Fig.4 are the measured electromechanical impedance spectrum of the 15.06×30.10×20.03mm$^3$ rectangular parallelepiped alpha-quartz sample from 50 to 310kHz at 25℃. It can be seen that although the measured conductance signals are relatively weaker (~0.1μS) than those in PZT sample (~10μS),[23] they are still almost noiseless. Below 150kHz, the resonance frequencies are far from each other. However, when the drive frequencies increase, the resonance modes are much denser. Due to the high-quality factors of alpha-quartz, the resonance curves are seldom overlapped, allowing the eigenmodes to be easily identified. In this work, 100 resonance modes are identified and used to extract the eight independent material coefficients. The photo of the quartz sample and the calculated 3-D deformations of the first four order resonance modes are shown in the top of Fig.4. The pink lines in Fig.4 represent the spectrum measured by the commercial sandwich-like RUS apparatus (Quasar RUSpec) using the same sample at the same temperature. It can be seen that the measured resonance frequencies are slightly larger than those by PE-EMIS. Fig.5 shows the relative errors of PE-EMIS and RUS for each resonance frequency. It can be seen that the maximum error can reach 0.3% and the average error is about 0.1%. These errors are caused by the clamping force in the sandwich-like RUS, which makes the corner of the sample become an elastic support condition (as shown in the insert of Fig.5) instead of a completely free boundary condition. According to Ohno, the degree of the shift of the resonance frequency in a clamped sample is not only related to the magnitude of the clamping force but also dependent on which pair of corners it is clamped.[16] So, it is absolutely impossible to eliminate this influence. In addition, the signals measured by the sandwich-like RUS may have some errors, for example, it is abnormal that the resonance peak is too wide between 150kHz and 155kHz (seen in Fig.4). Compared with that of the sandwich-like RUS, the spectrum determined by our proposed PE-EMIS is under completely free boundary condition and looks more reasonable, allowing the material coefficients to be extracted more accurately.

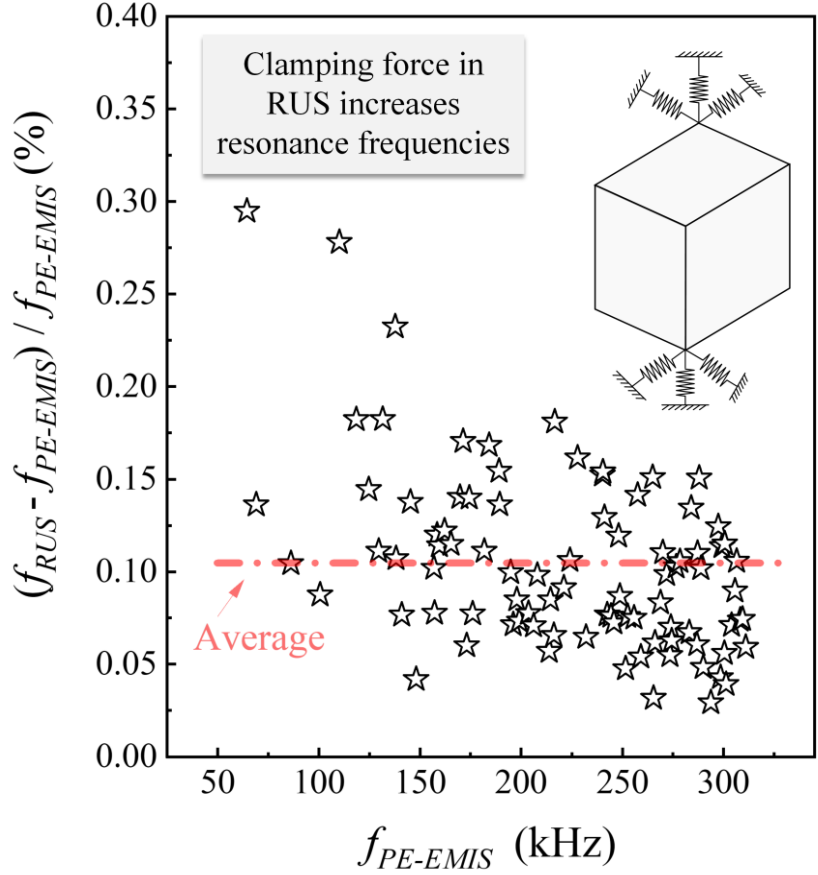

Fig.5 Relative measurement errors of PE-EMIS and sandwich-like RUS.

The independent elastic coefficients of alpha-quartz that we chose for inversion calculation are $c_{12}^E$, $c_{13}^E$, $c_{14}^E$, $c_{33}^E$, $c_{44}^E$, and $c_{66}^E$. Here, we use Legendre functions up to N=18, where $N = \lambda + \mu + \nu$. In this condition, the number of terms in the displacement and electric potential polynomials in Eqs. (5) and (6) are 3990 and 1330 respectively, which are sufficient for weak piezoelectric coefficients to be accurately extracted.[20] After paring up the measured frequencies (by PE-EMIS) with the calculated ones, 100 eigenmodes are identified and listed in Table I. Among them, eight eigenmodes are missing, which are Ag-2, Ag-4, Bu-11, Ag-13, Bg-9, Au-15, Bu-22 and Bg-23. The relative errors of measured and calculated resonance frequencies of PE-EMIS are also listed and almost all of them are less than 0.1%. The numerical calculation time for a single iteration step is about 50 seconds. We use the material coefficients in Ohno's paper[16] as the initial values, and 30 iterations are required to reach convergence. The inversion calculation using the measured spectrum of the sandwich-like RUS method is also conducted, and the measured and calculated resonance frequencies, as well as the relative errors, are listed in Table A in Appendix.

Table I Measured and calculated first 100 order resonance frequencies of the alpha-quartz sample by using PE-EMIS

| Mode | $f^{meas}$ (kHz) | $f^{cal}$ (kHz) | Diff(%) | Mode | $f^{meas}$ (kHz) | $f^{cal}$ (kHz) | Diff(%) |
|---|---|---|---|---|---|---|---|
| Au-1 | 54.99 | 54.93 | 0.10 | Bu-14 | 232.05 | 232.00 | 0.02 |
| Au-2 | 64.43 | 64.37 | 0.09 | Au-12 | 240.19 | 240.22 | -0.01 |
| Bu-1 | 69.05 | 68.96 | 0.13 | Bu-15 | 240.43 | 240.50 | -0.03 |
| Ag-1 | 86.15 | 86.06 | 0.11 | Bg-12 | 241.16 | 241.22 | -0.02 |
| Bg-1 | 100.56 | 100.67 | -0.11 | Ag-16 | 242.53 | 242.53 | 0.00 |
| Bg-2 | 110.09 | 110.12 | -0.02 | Bu-16 | 244.39 | 244.33 | 0.02 |
| Ag-2 | … | 114.75 | … | Au-13 | 245.83 | 245.82 | 0.01 |
| Au-3 | 118.52 | 118.61 | -0.08 | Ag-17 | 248.20 | 248.20 | 0.00 |
| Au-4 | 124.66 | 124.72 | -0.05 | Bg-13 | 248.84 | 248.81 | 0.01 |
| Bg-3 | 129.54 | 129.57 | -0.02 | Bu-17 | 251.60 | 251.55 | 0.02 |
| Ag-3 | 131.6 | 131.62 | -0.01 | Au-14 | 252.68 | 252.62 | 0.02 |
| Bg-4 | 137.68 | 137.90 | -0.16 | Bg-14 | 255.78 | 255.80 | -0.01 |
| Bu-2 | 138.10 | 138.08 | 0.02 | Ag-18 | 257.54 | 257.72 | -0.07 |
| Bg-5 | 141.00 | 140.97 | 0.02 | Bu-18 | 259.14 | 259.12 | 0.01 |
| Bu-3 | 145.28 | 145.24 | 0.03 | Bg-15 | 261.29 | 261.37 | -0.03 |
| Au-5 | 148.13 | 147.98 | 0.10 | Au-15 | … | 261.51 | … |
| Ag-4 | … | 151.42 | … | Au-16 | 264.98 | 265.14 | -0.06 |
| Au-6 | 156.86 | 156.84 | 0.02 | Ag-19 | 265.44 | 265.34 | 0.04 |
| Bu-4 | 157.23 | 157.08 | 0.10 | Bu-19 | 266.14 | 266.10 | 0.02 |
| Ag-5 | 158.47 | 158.38 | 0.05 | Bu-20 | 268.92 | 268.95 | -0.01 |
| Ag-6 | 159.55 | 159.73 | -0.11 | Au-17 | 269.97 | 270.07 | -0.04 |
| Bu-5 | 162.13 | 162.32 | -0.11 | Bg-16 | 271.94 | 272.06 | -0.05 |
| Bu-6 | 165.37 | 165.57 | -0.12 | Bu-21 | 273.59 | 273.46 | 0.05 |
| Au-7 | 169.73 | 169.76 | -0.02 | Au-18 | 274.02 | 273.87 | 0.05 |
| Ag-7 | 171.28 | 171.47 | -0.11 | Ag-20 | 274.46 | 274.34 | 0.04 |
| Ag-8 | 173.08 | 173.16 | -0.04 | Bg-17 | 276.56 | 276.47 | 0.03 |
| Bu-7 | 174.36 | 174.59 | -0.13 | Ag-21 | 278.53 | 278.51 | 0.01 |
| Bg-6 | 176.04 | 176.21 | -0.10 | Bu-22 | … | 281.08 | ... |
| Bu-8 | 181.75 | 181.92 | -0.09 | Bu-23 | 283.11 | 282.88 | 0.08 |
| Au-8 | 183.64 | 183.62 | 0.01 | Ag-22 | 283.99 | 284.07 | -0.03 |
| Ag-9 | 184.17 | 184.31 | -0.08 | Ag-23 | 286.89 | 286.83 | 0.02 |
| Bg-7 | 189.10 | 189.34 | -0.12 | Au-19 | 287.14 | 287.09 | 0.02 |
| Au-9 | 189.39 | 189.50 | -0.06 | Bg-18 | 288.01 | 288.20 | -0.07 |
| Ag-10 | 194.85 | 194.94 | -0.05 | Ag-24 | 288.87 | 288.58 | 0.10 |
| Bu-9 | 196.10 | 195.92 | 0.09 | Bg-19 | 289.98 | 289.87 | 0.04 |
| Bu-10 | 197.88 | 197.79 | 0.05 | Au-20 | 293.81 | 293.67 | 0.05 |
| Ag-11 | 199.07 | 199.09 | -0.01 | Bu-24 | 297.37 | 297.42 | -0.02 |
| Bg-8 | 203.65 | 203.58 | 0.04 | Bg-20 | 298.52 | 298.38 | 0.05 |
| Ag-12 | 206.17 | 206.06 | 0.05 | Au-21 | 299.30 | 299.33 | -0.01 |
| Au-10 | 207.96 | 207.88 | 0.04 | Ag-25 | 300.40 | 300.14 | 0.09 |
| Bu-11 | … | 208.02 | … | Ag-26 | 300.61 | 300.63 | -0.01 |
| Ag-13 | … | 211.08 | … | Bg-21 | 301.31 | 301.24 | 0.02 |
| Au-11 | 213.85 | 213.80 | 0.02 | Bu-25 | 304.00 | 303.92 | 0.03 |
| Bg-9 | … | 213.81 | … | Bg-22 | 305.63 | 305.54 | 0.03 |
| Bu-12 | 214.45 | 214.40 | 0.02 | Au-22 | 306.54 | 306.66 | -0.04 |

| | | | | | | |
|---|---|---|---|---|---|---|
| Bg-10 | 216.15 | 215.97 | 0.08 | Ag-27 | 308.26 | 308.22 | 0.01 |
| Bu-13 | 216.62 | 216.65 | -0.01 | Bg-23 | … | 309.07 | … |
| Ag-14 | 221.07 | 221.07 | 0.00 | Au-23 | 309.41 | 309.31 | 0.03 |
| Ag-15 | 224.09 | 224.09 | 0.00 | Ag-28 | 311 | 310.69 | 0.10 |
| Bg-11 | 227.84 | 227.83 | 0.00 | Bg-24 | 313.11 | 313.25 | -0.04 |

Table II Elastic coefficients $c_{ij}^E$ (GPa) and piezoelectric coefficients $e_{ij}$ (C/m$^2$) of alpha-quartz determined by PE-EMIS and sandwich-like RUS, as well as by other scholars using different methods.

| | PE-EMIS 25°C | RUS 25°C | $\left\|\frac{P_P-P_R}{P_P}\right\|$% | Ogi[20] 30°C | Kushibiki[10] 23°C | James[29] 25°C | Koga[9] 20°C | Bechmann[8] 20°C | $\left\|\frac{Max-Min}{Average}\right\|$% |
|---|---|---|---|---|---|---|---|---|---|
| $c_{11}^E$ | 86.72 | 86.79 | 0.08 | 86.76 | 86.80 | 86.79 | 86.83 | 86.74 | 0.13 |
| $c_{12}^E$ | 6.802 | 6.986 | 2.71 | 6.868 | 7.036 | 6.790 | 7.090 | 6.99 | 4.32 |
| $c_{13}^E$ | 11.81 | 11.75 | 0.51 | 11.85 | 11.94 | 12.01 | 11.93 | 11.91 | 2.19 |
| $c_{14}^E$ | -18.10 | -18.04 | 0.33 | -18.02 | -18.06 | -18.12 | -18.06 | -17.91 | 1.16 |
| $c_{33}^E$ | 105.22 | 105.49 | 0.26 | 105.46 | 105.78 | 105.79 | 105.94 | 107.2 | 1.87 |
| $c_{44}^E$ | 58.22 | 58.32 | 0.17 | 58.14 | 58.22 | 58.21 | 58.26 | 57.94 | 0.65 |
| $c_{66}^E$ | 39.96 | 39.90 | 0.15 | 39.95 | 39.88 | 40.00 | 39.87 | 39.88 | 0.33 |
| $e_{11}$ | 0.153 | 0.173 | 13.07 | 0.151 | 0.172 | 0.171 | 0.175 | 0.171 | 14.41 |
| $e_{14}$ | -0.040 | -0.070 | 75.00 | -0.061 | -0.039 | -0.041 | -0.041 | -0.041 | 65.17 |

The extracted elastic and piezoelectric coefficients of alpha-quartz by using PE-EMIS and sandwich-like RUS are presented in the 2nd and the 3rd columns of Table II. The material coefficients measured by other scholars using different methods are also listed, in the 5th-9th columns, for comparison. As shown in the 4th column of Table II, for PE-EMIS and sandwich-like RUS, the relative errors of the extracted elastic coefficients are very small (with the maximum error 2.71% and minimum error 0.08%). However, the relative errors for the piezoelectric coefficients $e_{11}$ and $e_{14}$ are very large and can reach 13% and 75%, respectively. The $e_{11}$ and $e_{14}$ extracted by sandwich-like RUS are the largest among the seven sets of data, and obviously, it is related to the increase of the resonance frequencies by the clamping force. According to the seven sets of data, we can also find that, the deviation (as shown in the last column of Table II) of $c_{12}^E$ and $c_{13}^E$ are obviously larger than the other elastic coefficients, and the deviation of $e_{11}$ and $e_{14}$ can reach 14% and 65%, respectively. To interpret this phenomenon, we calculate the normalized sensitivities of the elastic and piezoelectric coefficients of the alpha-quartz sample to the $k^{th}$ resonance frequency $f_k$ (i.e., $\frac{\partial f_k}{\partial c_{ij}^E}\frac{c_{ij}^E}{f_k}$ and $\frac{\partial f_k}{\partial e_{mn}}\frac{e_{mn}}{f_k}$) by using the extracted matrix coefficients from PE-EMIS test (the 2nd column of Table II). As can be seen from Fig.6, $c_{33}^E$, $c_{14}^E$, $c_{44}^E$, and $c_{66}^E$ are relatively sensitive to $f_k$, whereas the sensitivities of $c_{12}^E$ and $c_{13}^E$ to $f_k$ are much weaker. In addition, due to the extremely weak piezoelectricity of alpha-quartz, the normalized sensitivities of piezoelectric coefficients $e_{11}$ to $k^{\text{th}}$ resonance frequency are always less than 1%. For $e_{14}$ the normalized sensitives are even less than 0.15%. It should be note that, in PZT ceramic sample, these values can reach 10% or even 20%.[23] So, for alpha-quartz sample, small resonance frequency measurement errors can induce large deviations in the inversion results of $e_{11}$ and $e_{14}$, and this is why the extracted values using the spectrum measured by sandwich-like RUS are so large. By varying the sample's dimensions, the same calculations are carried out, and we find that the normalized sensitivities of $c_{12}^E$, $c_{13}^E$, $e_{11}$ and $e_{14}$ to $f_k$ are always very small, and this is why the deviations of these measured coefficients in Table II are large than others.

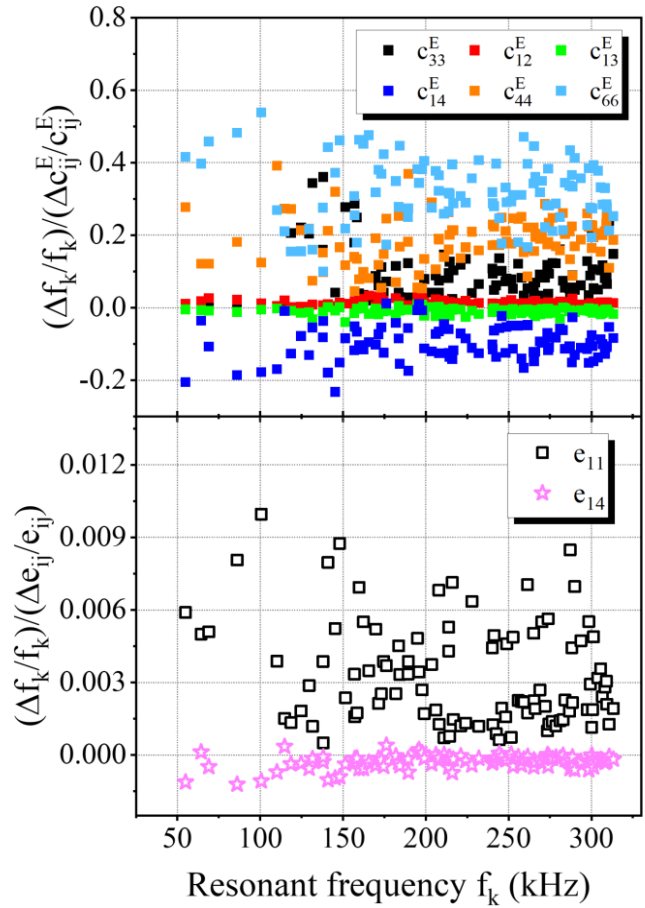

Fig.6 Normalized sensitivities of the elastic and piezoelectric coefficients to the $k^{\text{th}}$ resonance frequency of the 15.06×30.10×20.03mm³ alpha-quartz.

## 5.2 Mode identification using off-plane displacement measurement

To prove each resonance mode in PE-EMIS test to be unambiguously identified and to guarantee all of the material coefficients to be accurately determined, we followed Ogi *et al.*[20, 21, 30] to use a scanned-laser vibrometer to determine the off-plane displacement distributions of the $x_3$ face (15.06×30.10mm$^2$) of the quartz sample under different resonance modes and compare with the calculated ones. Due to the transparency of alpha-quartz crystal, a non-conductive reflective paint is sprayed on one of the $x_3$ face of the quartz sample before testing, and the paint thickness is about 1μm. During the test, the sample is excited at the fixed resonance frequency using the same impedance analyzer. The single step distance of scanned laser along $x_1$ and $x_2$ directions are 1mm and 2mm, respectively. Fig.7 shows the measured and calculated results for resonance modes Bg-5, Bu-3, Ag-10 and Au-10, respectively. It can be seen that the measured patterns coincide with the calculated ones very well.

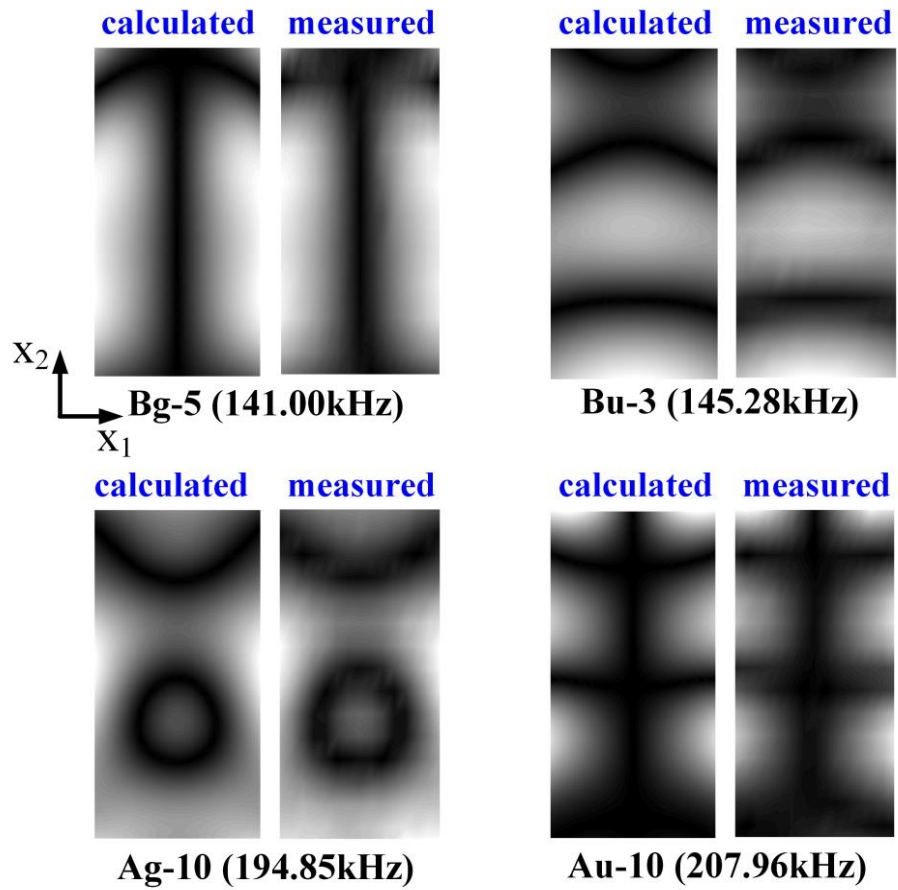

Fig.7 Calculated and measured off-plane displacement distribution of the x3-face of the alpha-quartz sample under resonance modes Bg-5, Bu-3, Ag-10 and Au-10.

## 5.3 Internal frictions

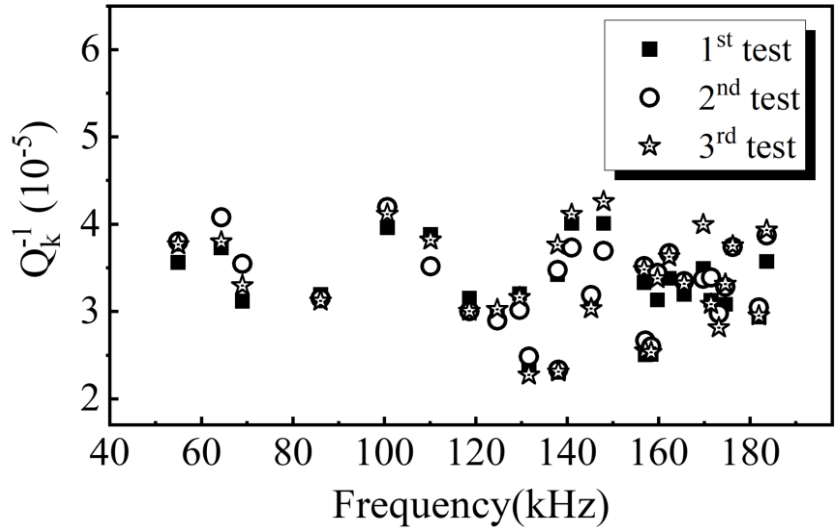

Fig. 8 1st, 2nd, and 3rd measured total internal friction $Q_k^{-1}$ associated with the first thirty resonance frequencies (excluding two missing modes Ag-2 and Ag-4).

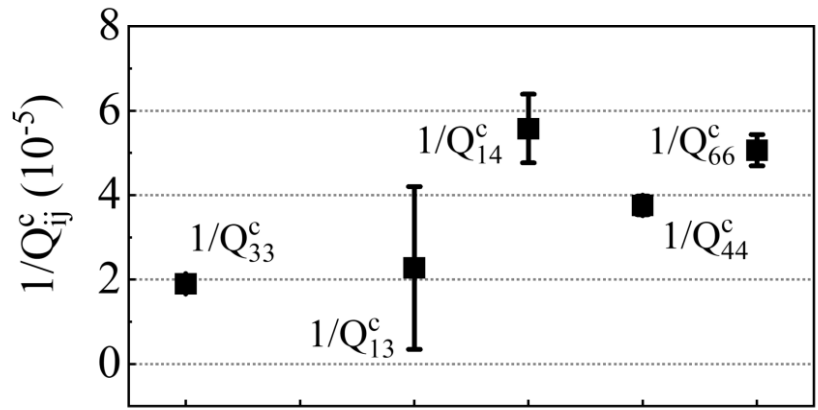

Fig. 9 Calculated independent elastic internal frictions $\frac{1}{Q_{ij}^c}$ of alpha-quartz.

As mentioned above, for elastic and piezoelectric coefficients measurement, the quartz sample can be directly supported by a soft foam with very low acoustic impedance, and the complete spectrum can then be conveniently detected with an impedance analyzer in one test. However, in this condition, the measured internal friction of k$^{th}$ resonance mode $Q_k^{-1}$ is much larger than its real values. This is due to that the vibration energy is transmitted into the soft foam, causing the additional energy loss. For piezoelectricity with large internal frictions, such as PZT ceramic (Q<1000), the support manner has little effect on the measured internal frictions.[23] However, for the 'lossless' materials, such as alpha-quartz, we must use needles to support it near the displacement nodes of each eigenmode to avoid additional energy loss, as shown in Fig.3. Then, the internal friction of k$^{th}$ eigenmode $Q_k^{-1}$ can be relatively accurately determined one by one with the impedance analyzer. Fig.8 shows the measured total internal frictions $Q_k^{-1}$ associated with the first thirty resonance frequencies (excluding two missing modes Ag-2 and Ag-4). The tests are carried out three times, and are denoted by 1$^{st}$, 2$^{nd}$ and 3$^{rd}$. The relative errors of the three measured $Q_k^{-1}$ are typically less than 12%. According to Eqs. (20) and (21), we only need to consider the elastic internal frictions $\frac{1}{Q_{12}^c}$, $\frac{1}{Q_{13}^c}$, $\frac{1}{Q_{14}^c}$, $\frac{1}{Q_{33}^c}$, $\frac{1}{Q_{44}^c}$, and $\frac{1}{Q_{66}^c}$. Assume that all the elastic internal frictions are frequency-independent during the testing frequency range. Based on the measured $Q_k^{-1}$, the $\frac{1}{Q_{ij}^c}$ can be calculated using the least square method, and the results are shown in Fig.9. It can be seen that, the measured values of $\frac{1}{Q_{33}^c}$, $\frac{1}{Q_{44}^c}$ and $\frac{1}{Q_{66}^c}$ are relative stable (with error less than $3\times10^{-6}$). The shear internal frictions $\frac{1}{Q_{44}^c}$ and $\frac{1}{Q_{66}^c}$ exceed the longitudinal internal frictions $\frac{1}{Q_{33}^c}$ by a factor of about 2-2.5, which is similar as that in metals and PZT ferroelectric ceramic.[23] The calculated $\frac{1}{Q_{12}^c}$ and $\frac{1}{Q_{13}^c}$ are not accurate and is due that the sensitivities of $c_{12}^E$ and $c_{13}^E$ to $f_k$ are extremely weak (as seen in Fig.6). The calculated $\frac{1}{Q_{12}^c}$ is negative, so it is not presented in Fig.9. The negative $\frac{1}{Q_{ij}^c}$ is also reported by Sumino in the measurement of single MgO.[18] Although the measured $\frac{1}{Q_{12}^c}$ and $\frac{1}{Q_{13}^c}$ is not accurate, the measured $\frac{1}{Q_{33}^c}$, $\frac{1}{Q_{44}^c}$ and $\frac{1}{Q_{66}^c}$ are very reliable and can be used in practice.

## 6. Conclusion

In summary, we employed our recently proposed partial-electrode electromechanical impedance spectroscopy (PE-EMIS) to determine the independent elastic coefficients, piezoelectric coefficients, and internal frictions of alpha-quartz crystal. The first 100 order eigenmodes measured by an impedance analyzer are noiseless and under perfectly free boundary conditions, allowing precise extraction of material coefficients. To avoid mode misidentification, the off-plane displacement distributions of different resonance modes of the quartz sample are also detected with the scanned laser vibrometer. For comparison, the commercial sandwich-like RUS method is also utilized and the results reveal that the elastic coefficients determined by these two methods agree well with each other and with those determined by other scholars using various techniques. However, the piezoelectric coefficients $e_{11}$ and $e_{14}$ extracted from sandwich-like RUS are visibly inaccurate, demonstrating that clamping force has large influence. In comparison to the sandwich-like RUS, our proposed PE-EMIS is more effective and convenient, and will be extensively utilized for accurately characterization of piezoelectric crystals.

## Appendix

Table A Measured and calculated first 100 order resonance frequencies of the alpha-quartz sample by using sandwich-like RUS

| Mode | $f^{meas}$ (kHz) | $f^{cal}$ (kHz) | Diff(%) | Mode | $f^{meas}$ (kHz) | $f^{cal}$ (kHz) | Diff(%) |
|---|---|---|---|---|---|---|---|
| Au-1 | 54.97 | 54.98 | -0.02 | Bu-14 | 232.20 | 232.18 | 0.01 |
| Au-2 | 64.62 | 64.44 | 0.28 | Au-12 | 240.56 | 240.50 | 0.02 |
| Bu-1 | 69.14 | 69.03 | 0.17 | Bu-15 | 240.80 | 240.71 | 0.04 |
| Ag-1 | 86.24 | 86.15 | 0.11 | Bg-12 | 241.47 | 241.50 | -0.01 |
| Bg-1 | 100.65 | 100.76 | -0.11 | Ag-16 | 242.72 | 242.70 | 0.01 |
| Bg-2 | 110.40 | 110.24 | 0.14 | Bu-16 | 244.58 | 244.56 | 0.01 |
| Ag-2 | 114.92 | 114.90 | 0.02 | Au-13 | 246.01 | 246.09 | -0.03 |
| Au-3 | 118.74 | 118.76 | -0.02 | Ag-17 | 248.50 | 248.46 | 0.02 |
| Au-4 | 124.84 | 124.86 | -0.02 | Bg-13 | 249.06 | 249.11 | -0.02 |
| Bg-3 | 129.68 | 129.72 | -0.03 | Bu-17 | 251.72 | 251.81 | -0.03 |
| Ag-3 | 131.84 | 131.80 | 0.03 | Au-14 | 252.87 | 252.87 | 0.00 |
| Bg-4 | 138.00 | 138.04 | -0.03 | Bg-14 | 255.97 | 256.12 | -0.06 |
| Bu-2 | 138.25 | 138.28 | -0.02 | Ag-18 | 257.90 | 257.98 | -0.03 |
| Bg-5 | 141.11 | 141.10 | 0.01 | Bu-18 | 259.28 | 259.32 | -0.01 |
| Bu-3 | 145.48 | 145.40 | 0.05 | Bg-15 | 0.00 | 261.64 | … |
| Au-5 | 148.19 | 148.13 | 0.04 | Au-15 | 0.00 | 261.82 | … |
| Ag-4 | 151.75 | 151.64 | 0.07 | Au-16 | 265.38 | 265.45 | -0.03 |
| Au-6 | 157.02 | 157.00 | 0.01 | Ag-19 | 265.52 | 265.56 | -0.01 |
| Bu-4 | 157.35 | 157.31 | 0.03 | Bu-19 | 266.30 | 266.39 | -0.03 |
| Ag-5 | 158.66 | 158.57 | 0.06 | Bu-20 | 269.14 | 269.23 | -0.03 |
| Ag-6 | 159.73 | 159.88 | -0.09 | Au-17 | 270.27 | 270.37 | -0.04 |
| Bu-5 | 162.33 | 162.46 | -0.08 | Bg-16 | 272.21 | 272.38 | -0.06 |
| Bu-6 | 165.56 | 165.73 | -0.10 | Bu-21 | 273.74 | 273.73 | 0.00 |
| Au-7 | 169.97 | 169.92 | 0.03 | Au-18 | 274.21 | 274.16 | 0.02 |
| Ag-7 | 171.57 | 171.68 | -0.06 | Ag-20 | 274.63 | 274.63 | 0.00 |
| Ag-8 | 173.18 | 173.32 | -0.08 | Bg-17 | 276.84 | 276.81 | 0.01 |
| Bu-7 | 174.60 | 174.74 | -0.08 | Ag-21 | 278.82 | 278.75 | 0.03 |
| Bg-6 | 176.18 | 176.44 | -0.15 | Bu-22 | 281.64 | 281.43 | 0.08 |
| Bu-8 | 181.95 | 182.13 | -0.10 | Bu-23 | 283.30 | 283.18 | 0.04 |
| Au-8 | … | 183.80 | … | Ag-22 | 284.37 | 284.31 | 0.02 |
| Ag-9 | 184.48 | 184.46 | 0.01 | Ag-23 | 287.06 | 287.09 | -0.01 |
| Bg-7 | 189.39 | 189.52 | -0.07 | Au-19 | 287.46 | 287.39 | 0.02 |
| Au-9 | 189.65 | 189.72 | -0.04 | Bg-18 | 288.44 | 288.50 | -0.02 |
| Ag-10 | 195.04 | 195.17 | -0.06 | Ag-24 | 289.16 | 288.84 | 0.11 |
| Bu-9 | 196.24 | 196.11 | 0.07 | Bg-19 | 290.12 | 290.16 | -0.02 |
| Bu-10 | 198.05 | 197.99 | 0.03 | Au-20 | 293.90 | 293.98 | -0.03 |
| Ag-11 | 199.22 | 199.26 | -0.02 | Bu-24 | 297.74 | 297.76 | -0.01 |
| Bg-8 | 203.81 | 203.81 | 0.00 | Bg-20 | 298.65 | 298.67 | -0.01 |
| Ag-12 | 206.32 | 206.29 | 0.01 | Au-21 | 299.64 | 299.68 | -0.01 |
| Au-10 | 208.16 | 208.11 | 0.03 | Ag-25 | 300.57 | 300.48 | 0.03 |
| Bu-11 | … | 208.18 | … | Ag-26 | 300.96 | 300.97 | 0.00 |
| Ag-13 | 211.49 | 211.32 | 0.08 | Bg-21 | 301.43 | 301.51 | -0.03 |
| Au-11 | 213.97 | 214.04 | -0.03 | Bu-25 | 304.22 | 304.22 | 0.00 |
| Bg-9 | 214.06 | 214.06 | 0.00 | Bg-22 | 305.90 | 305.90 | 0.00 |
| Bu-12 | 214.63 | 214.67 | -0.02 | Au-22 | 306.86 | 307.03 | -0.05 |
| Bg-10 | 216.29 | 216.20 | 0.04 | Ag-27 | 308.49 | 308.52 | -0.01 |
| Bu-13 | 217.01 | 216.88 | 0.06 | Bg-23 | 309.34 | 309.43 | -0.03 |
| Ag-14 | 221.27 | 221.26 | 0.00 | Au-23 | 309.64 | 309.66 | -0.01 |
| Ag-15 | 224.33 | 224.35 | -0.01 | Ag-28 | 311.18 | 311.02 | 0.05 |
| Bg-11 | 228.21 | 228.10 | 0.05 | Bg-24 | 313.46 | 313.62 | -0.05 |

## Declaration of Competing Interest

The authors declare that they have no known competing financial interests or personal relationships that could have appeared to influence the work reported in this paper.

**Data Availability**

The data that support the findings of this study are available from the corresponding author upon reasonable request

**Acknowledgement**

We thank Dr. Hengtong Bu and Prof. Yang Shao in Tsinghua University for providing commercial RUSepc test.

This work is supported by the National Natural Science Foundation of China under Grant No. 12172007.

**References**

[1] R.L. Bunde, E.J. Jarvi, J.J. Rosentreter, Piezoelectric quartz crystal biosensors, Talanta, 46 (1998) 1223-1236.
[2] V.S. Kumar, V. Priya, Recent Advances in Quartz Crystal Microbalance-Based Sensors, Journal of Sensors,2011,(2011-9-19), 2011 (2011) 539-556.
[3] V.E. Granstaff, S.J. Martin, Characterization of a thickness-shear mode quartz resonator with multiple nonpiezoelectric layers, Journal of Applied Physics, 75 (1994) 1319-1329.
[4] Brice, C. J., Crystals for quartz resonators, Rev Modern Phys, 57 (1985) 105-146.
[5] Mason, P. W., Electrical wave filters employing quartz crystals as elements, Bell System Technical Journal, 13 (2014) 405-452.
[6] Tancrell, H. R., Holland, G. M., Acoustic surface wave filters, Proceedings of the IEEE, (1971).
[7] P.J. Heaney, D.R. Veblen, Observationsof the alpha-beta phase transition in quartz: A review of imaging and diffraction studies and some new results, American Mineralogist, 76 (1991) 1018-1032.
[8] Bechmann, R., Elastic and Piezoelectric Constants of Alpha-Quartz, Phys Rev, 110 (1958) 1060-1061.
[9] I. Koga, M. Aruga, Y. Yoshinaka, Theory of Plane Elastic Waves in a Piezoelectric Crystalline Medium and Determination of Elastic and Piezoelectric Constants of Quartz, Phys Rev, 109 (1958) 1467-1473.
[10] J.-i. Kushibiki, M. Ohtagawa, I. Takanaga, Comparison of acoustic properties between natural and synthetic α-quartz crystals, Journal of Applied Physics, 94 (2003) 295-300.
[11] V.Y. Topolov, C.R. Bowen, Inconsistencies of the complete sets of electromechanical constants of relaxor-ferroelectric single crystals, Journal of Applied

Physics, 109 (2011) 27.

[12] J. Maynard, Resonant ultrasound spectroscopy, Physics Today, 49 (1996) 26-31.

[13] A. Migliori, J.D. Maynard, Implementation of a modern resonant ultrasound spectroscopy system for the measurement of the elastic moduli of small solid specimens, Review of Scientific Instruments, 76 (2005) 1113.

[14] B.J. Zadler, J. Rousseau, J.A. Scales, M.L. Smith, Resonant Ultrasound Spectroscopy: theory and application, Geophysical Journal of the Royal Astronomical Society, 156 (2010) 154-169.

[15] A. Migliori, T.W. Darling, Resonant ultrasound spectroscopy for materials studies and non-destructive testing, Ultrasonics, 34 (1996) 473-476.

[16] I. Ohno, Rectangular parallellepiped resonance method for piezoelectric crystals and elastic constants of alpha-quartz, Physics and Chemistry of Minerals, 17 (1990) 371-378.

[17] H. Guo, A. Lal, Investigation of clamping force in resonant ultrasound spectroscopy, The Journal of the Acoustical Society of America, 112 (2002) 2439-2439.

[18] Y. Sumino, I. Ohno, T. Goto, M. Kumazawa, MEASUREMENT OF ELASTIC CONSTANTS AND INTERNAL FRICTIONS ON SINGLE-CRYSTAL MgO BY RECTANGULAR PARALLELEPIPED RESONANCE, Journal of Physics of the Earth, 24 (1976) 263-273.

[19] R.G. Leisure, K. Foster, J.E. Hightower, D.S. Agosta, Internal friction studies by resonant ultrasound spectroscopy, Materials Science & Engineering A, 370 (2004) 34-40.

[20] H. Ogi, T. Ohmori, N. Nakamura, M. Hirao, Elastic, anelastic, and piezoelectric coefficients of α-quartz determined by resonance ultrasound spectroscopy, Journal of Applied Physics, 100 (2006) 299.

[21] H. Ogi, N. Nakamura, K. Sato, M. Hirao, S. Uda, Elastic, anelastic, and piezoelectric coefficients of langasite: resonance ultrasound spectroscopy with laser-Doppler interferometry, IEEE Transactions on Ultrasonics, Ferroelectrics, and Frequency Control, 50 (2003) 553-560.

[22] W.L. Johnson, S.A. Kim, S. Uda, Acoustic loss in langasite and langanite, in: IEEE International Frequency Control Symposium and PDA Exhibition Jointly with the 17th European Frequency and Time Forum, 2003. Proceedings of the 2003, 2003, pp. 646-649.

[23] M. Xie, F. Li, Determining Full Matrix Constants of Piezoelectric Crystal From a Single Sample Using Partial Electrode Electromechanical Impedance Spectroscopy, IEEE Transactions on Ultrasonics, Ferroelectrics, and Frequency Control, 69 (2022) 2984-2994.

[24] J. Pujol, The solution of nonlinear inverse problems and the Levenberg-Marquardt method, GEOPHYSICS, 72 (2007) W1-W16.

[25] R. Holland, E.P. EerNisse, Variational Evaluation of Admittances of Multielectroded Three-Dimensional Piezoelectric Structures, IEEE Transactions on Sonics and Ultrasonics, 15 (1968) 119-131.

[26] E. Mochizuki, APPLICATION OF GROUP THEORY TO FREE

OSCILLATIONS OF AN ANISOTROPIC RECTANGULAR PARALLELEPIPED, Journal of Physics of the Earth, 35 (1987) 159-170.

[27] M. Xie, F. Li, A modified piezoelectric ultrasonic composite oscillator technique for simultaneous measurement of elastic moduli and internal frictions at varied temperature, Review of Scientific Instruments, 91 (2020) 015110.

[28] M. Xie, F. Li, New method enables multifunctional measurement of elastic moduli and internal frictions, Journal of Applied Physics, 128 (2020) 230902.

[29] B.J. James, A new measurement of the basic elastic and dielectric constants of quartz, in: Proceedings of the 42nd Annual Frequency Control Symposium, 1988., 1988, pp. 146-154.

[30] H. Ogi, K. Sato, T. Asada, M. Hirao, Complete mode identification for resonance ultrasound spectroscopy, The Journal of the Acoustical Society of America, 112 (2002) 2553-2557.